\documentclass[conference]{IEEEtran}
\IEEEoverridecommandlockouts
\usepackage{cite}
\usepackage{amsmath,amssymb,amsfonts}
\usepackage{graphicx}
\usepackage{textcomp}
\usepackage{xcolor}
\usepackage{multirow}
\usepackage{hyperref}
\usepackage{algorithm}
\usepackage{algpseudocode}
\def\BibTeX{{\rm B\kern-.05em{\sc i\kern-.025em b}\kern-.08em
    T\kern-.1667em\lower.7ex\hbox{E}\kern-.125emX}}
\begin{document}

\title{MalDicom: A Memory Forensic Framework for Detecting Malicious Payload in DICOM Files\\
}

\author{\IEEEauthorblockN{\textsuperscript{} Ayushi Mishra}
\IEEEauthorblockA{\textit{Computer Science and Engineering} \\
\textit{Indian Institute of Technology}\\
Kanpur, India \\
ayushim@cse.iitk.ac.in}
\and
\and
\IEEEauthorblockN{\textsuperscript{} Priyanka Bagade}
\IEEEauthorblockA{\textit{Computer Science and Engineering} \\
\textit{Indian Institute of Technology}\\
Kanpur, India \\
pbagade@cse.iitk.ac.in}
}

\maketitle

\begin{abstract}

Digital Imaging and Communication System (DICOM) is widely used throughout the public health sector for portability in medical imaging. However, these DICOM files have vulnerabilities present in the preamble section. Successful exploitation of these vulnerabilities can allow attackers to embed executable codes in the 128-Byte preamble of DICOM files. Embedding the malicious executable will not interfere with the readability or functionality of DICOM imagery. However, it will affect the underline system silently upon viewing these files. This paper shows the infiltration of Windows malware executables into DICOM files. On viewing the files, the malicious DICOM will get executed and eventually infect the entire hospital network through the radiologist's workstation. The code injection process of executing malware in DICOM files affects the hospital networks and workstations' memory. Memory forensics for the infected radiologist's workstation is crucial as it can detect which malware disrupts the hospital environment, and future detection methods can be deployed. In this paper, we consider the machine learning (ML) algorithms to conduct memory forensics on three memory dump categories: Trojan, Spyware, and Ransomware, taken from the CIC-MalMem-2022 dataset. We obtain the highest accuracy of 75\% with the Random Forest model. For estimating the feature importance for ML model prediction, we leveraged the concept of Shapley values. 

\end{abstract}


\section{Introduction}

In medical imaging, Digital Imaging and Communication in Medicine (DICOM) file\cite{imaging_device} is the most commonly used file format, shared across hospital networks. It is a mainstream conduit for storing digital medical images such as X-rays, CT scans, and MRIs. These files hold sensitive information about patient data like name, date of birth, age, security number, medical history, and prescription record. Although the confidentiality of DICOM files complies with the Health Insurance and Portability Act (HIPAA), these files can be hacked for patient data and disrupt the hospital network\cite{PACS_cybersecurity}. At the same time, DICOM includes provisions for security and encryption. However, the decisions to encrypt these files are rarely implemented in practice\cite{security_dicom} due to the following probable reasons:
\begin{itemize}
    \item \textbf{Interoperability and Compatibility:} An encryption algorithm adds an extra layer of complexity and may interfere with the ability to view the DICOM files across different systems and software.
    \item \textbf{Performance and Speed:} Medical images like CT scans and MRIs can be massive. Adding encryption will increase the processing overhead. Quick access to medical images may precede encryption in time-sensitive emergencies.
    \item \textbf{Accessible Collaboration:} Healthcare professionals from different organizations may need to collaborate and share medical images in specific scenarios. Encrypting the DICOM files can make accessing these images more challenging without proper decryption keys, potentially hindering collaboration efforts.
\end{itemize}

DICOM files are enormous as they contain a vast amount of information. The empty space in the 128-Byte preamble of the DICOM header can allow the attackers to hide the malicious payload\cite{DICOM_vulnerability} in it without getting it noticed by medical professionals. This way, malware can hide behind the HIPAA-compliant DICOM files. These compromised DICOM files can disrupt operations, perform malicious manipulation of medical images, and can get unauthorized access to the system's resources and sensitive information. As per our knowledge, the vulnerability with the 128-Byte preamble of the DICOM file was never exploited before with the code injection attack. Although many security researchers emphasized the vulnerability present in the DICOM files\cite{DICOM_vulnerability}\cite{DICOM_vulnerability_2}, in this paper, we tried to answer the question: \emph{"Can DICOM vulnerabilities be exploited by implementing the code injection method to infiltrate malware in medical images and detect it using memory forensic technique?"} The paper shows the code injection process of malicious payload in DICOM files taking advantage of the empty 128-byte preamble section and eventually compromising the hospital network on its execution.


The malicious payload inserted in DICOM can affect the memory of the underline system. Such vulnerabilities will disrupt the IT systems and platforms such as Picture Archiving and Communication Systems (PACS)\cite{PACS}. The PACS is a piece of integrated medical technology that healthcare professionals use to manage, store, and retrieve digital images. All the digital scans from doctors, such as CT and MRI, are stored in the central repository, allowing healthcare providers to access the medical image. During the insertion of malware into the DICOM images, these infected medical images are stored inside the PACS server. The infiltration of malicious code can reside in the PACS system's volatile memory, making it crucial to investigate the system's RAM and study the impact of the malware on memory. The memory-resident malware that causes infection through code injection can affect the system's integrity. These malicious payloads will be written directly on the memory. This makes memory forensics an essential approach to analyze the infected memory and learn the execution path of malware to devise strategies and detect malware before it disrupts the hospital network in the future. To fulfill these investigation requirements, we propose a memory forensics framework MalDicom that uses ML algorithms to analyze the affected memory dumps and detect the malware class. We leveraged the Shapley values to validate and explain the malware classification results produced by ML algorithms mathematically.
The main contributions of this paper are enlisted as follows:

\begin{itemize}
    \item  Novel attack on DICOM files by inserting a malicious payload using the code injection attack. 
    \begin{itemize}
        \item  Eavesdrop in the network using a man-in-the-middle (MITM) attack to get the legitimate DICOM file, modify it and insert it back into the network.
        \item Inject the malicious payload into the 128-Byte preamble section of the DICOM file.
    \end{itemize}
    \item Development of memory forensic framework, MalDicom, to investigate the memory dump obtained from the infected machine after the execution of the malicious DICOM files.
    \begin{itemize}
        \item Use ML algorithms to classify the malware from the memory dump.
    \end{itemize}
    \item Use Shapley values for explainability of the ML model predictions to show which features have more impact on the output.
\end{itemize}

The rest of the paper is organized as follows: Section~\ref{related_work} describes the work done in the malware detection and memory forensics field. The attack scenario is explained in section~\ref{experimental section}. Within this section, we explained the process of infiltrating malicious payload in DICOM. After creating attacks, the investigation process using the proposed MalDicom framework is discussed in section~\ref{investigation}, where we leveraged ML algorithms to detect the malicious memory dump. We introduced the concept of Shapley values in subsection~\ref{feature} to predict which features play more importance in the model output. Section~\ref{results1} explains the results in the paper. Future work is discussed in section~\ref{future work}. We concluded the paper in section~\ref{conclusion}.  

\begin{figure*}[ht]
\centering
\includegraphics[scale=0.5]{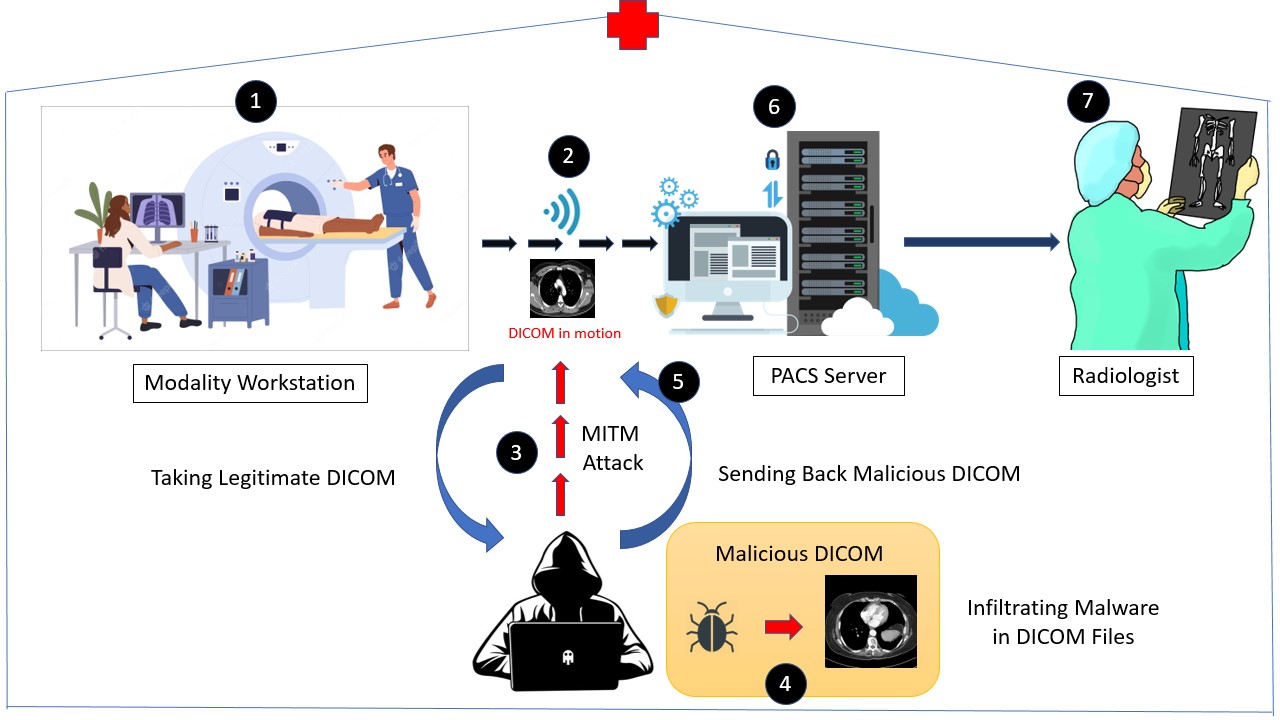}
\caption{An Attack Scenario in Hospital Network is described. The CT/MRI scans, when sent from the modality station to the PACS server, the MITM attack can steal the DICOM files in motion. The Code Injection is performed on DICOM and malicious files are sent back to the PACS server. Finally, the radiologist's workstation receives the infected DICOM files.}
\label{fig:1}
\end{figure*}

\section{Related Work}
\label{related_work}
Till now, limited research has been done on DICOM security. Researchers have expressed the severity of hiding malware inside the DICOM by sending a malicious payload in the doctor's mail and infecting the workstation\cite{dicom_security}. This is a classic malspam example. Few works focus on the security of the PACS server over the public network\cite{pacs_security}\cite{imaging_security}. They encouraged ensuring encryption of the DICOM file, digital signatures, and watermarking techniques. Mirsky et al.\cite{CT_GAN} show how an attacker can use an interceptor to perform a man-in-the-middle attack in the hospital by using generative adversarial network (GAN) models to inject and remove the lung cancer nodules from the patient's DICOM files for CT scans. Velinov et al.\cite{Dicom_hacked} presented their work POSTER in which they launched a ZIP bomb attack against the DICOM-enabled devices. The attack caused the device to run out of memory. However, these attacks have yet to be performed on DICOM files. 

The previous attack on DICOM files involved modifying the content of the file data. The code injection attacks on DICOM will not modify the file data to keep the readability and functionality of the DICOM same as before. They can silently affect the image-viewing workstation which makes the attack undetectable using currently available methods. Code injection attacks have been well-studied for SQL injection\cite{sql_injection} and Cross-Site Scripting (XSS)\cite{xss}. In this paper, we show how code injection attacks can be performed by inserting malware in DICOM files without getting noticed by medical practitioners.

Most of the malware resides inside the memory. Therefore, memory forensic techniques can help analyze the bad process and capture evidence of the behavior and source of the malware. Bajpai et al.\cite{ransomware} proposed a physical memory forensics method to facilitate complete data recovery for ransomware malware. Another system built for memory retrieval, called DroidScraper was developed by Ali-Gombe et al.\cite{droid}. This system can recover vital objects from allocated memory regions. This memory analysis technique targets the new Andriod Runtime (ART). The memory images obtained after collecting the artifacts are an important source of information about malware capabilities. Alrawi et al.\cite{forecast} create a probabilistic model for forecasting the capabilities of real-world malware and futuristic attacks. Each capability is also weighted according to the relative likelihood of execution.

Most of the research on memory forensics has been restricted to computer systems. However, with the growth of ubiquitous computing and the Medical Internet of Things (MIoT), the investigation against cyber attacks on medical devices can provide valuable evidence to detect life-threatening attacks on hospital networks. Mishra et al.\cite{MIoT_forensics} developed an innovative methodology combining an intrusion detection system and physiological data modeling to create a MIoT system suitable for forensic use. V.Schmitt\cite{med_forensics} has focused on how to build a more secure and forensic-ready medical device. As the health industry moves to Industry 4.0, practitioners need to learn and determine about digital forensics and incident response (DFIR)\cite{dfir} when dealing with medical devices. To this effect, the proposed MalDicom framework can be used to investigate attacks on medical devices. 

In this paper, we presented the memory forensic investigation framework, MalDicom on the memory dumps for malware inserted in DICOM files. We describe an attack scenario of how a DICOM can be hacked from the hospital networks, related vulnerabilities, and the infiltration of malware executables in section~\ref{attack}.

\section{Experimental Setup}
\label{experimental section}

\subsection{Attack Scenario}
\label{attack}

Figure ~\ref{fig:1} explains the attack scenario in the hospital network where an internal attack can lead to stealing DICOM files. A DICOM file in the hospital traverses through several stations, which increases its chance of getting infected. The CT scan or the MRI machine in the hospitals serves as the modality workstation where the images are stored in the \emph{.dcm} format, the standard medical image protocol. When in transit, these DICOM files are transmitted to the PACS\cite{PACS} machine. A PACS is an ethernet-based network where a centralized server collects scans from an attached CT/MRI modality. These files can be stored in the database and later retrieved for radiologists to analyze and annotate. Assuming the PACS server is available inside the hospitals, these DICOM files can be intercepted using a Man-in-the-Middle attack (MITM).

In this condition, the attacker can take the legitimate DICOM files, infiltrate malware inside it and then send back the malicious DICOM to the PACS Server. The PACS server then transfers these malicious files to the radiologist's workstation. In contrast, when the radiologist opens the CT/ MRI scans of the patients, the malware gets executed, and ultimately the entire hospital network is compromised. Again, we asked the question: \emph{"How can a code injection attack affect the DICOM viewer?"}. Malicious DICOM files with code injection attacks may exploit vulnerabilities in the software or systems used for viewing or processing these files. By injecting malicious code, an attacker could exploit the software's security flaws to gain unauthorized access, execute arbitrary commands, or escalate privileges. Code injection attacks can be utilized as an entry point for delivering malware into the DICOM viewer or the systems it interacts with. The attacker may inject malicious code that downloads and executes additional malware, such as ransomware, keyloggers, or remote access trojans (RATs). This can further compromise the security and privacy of patient data and the entire hospital network. 

\subsection{MITM Attack Setup}

\begin{figure}[ht]
\centering
\includegraphics[width=0.5\textwidth]{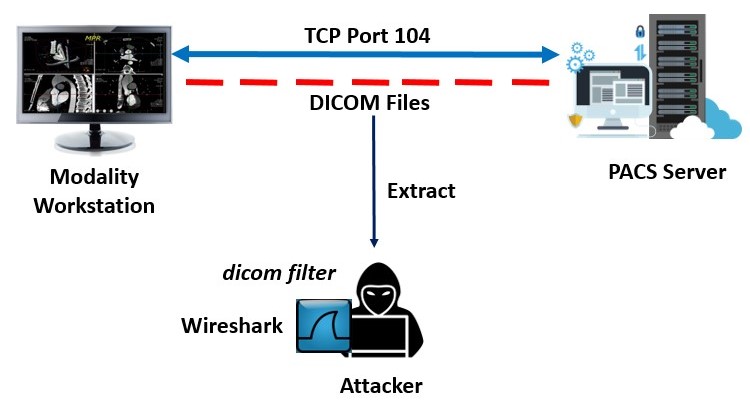}
\caption{Man-in-the-Middle Attack for extracting DICOM files}
\label{fig:mitm}
\end{figure}

We have used three machines to execute the MITM attack, a Philips PC with Ubuntu OS as an attacker, a Lenovo Yoga laptop with Windows 11 OS as a modality workstation, and a Dell laptop as a PACS server, used for building an authentic connection with the workstation. We have chosen a Windows modality workstation as most hospitals still use old legacy systems that rely on Windows OS. Ettercap\cite{ettercap}, an open-source tool, is used to create a MITM attack. The PACS server and the modality workstation communicated with a WLAN connection. The attacker's machine had Wireshark\cite{wireshark} tool running to eavesdrop on the network traffic. In a typical scenario where DICOM files are transmitted over a TCP connection, the DICOM data is divided into packets and encapsulated within TCP segments. Wireshark can capture these TCP segments, providing a detailed view of the network traffic\cite{network_forensics}. To isolate DICOM traffic, the captured packets can be filtered using Wireshark's display filter $dicom$. Wireshark provides filters focusing on specific protocols, ports, or DICOM-specific fields within the packet. The figure~\ref{fig:mitm} describes a MITM attack showing how the DICOM files can be analyzed and extracted. The attacker uses the Ettercap tool to capture the username and password of the targeted machine. Once the machine's confidentiality is hampered, the attacker can easily extract the DICOM files, manipulate them with the malware, and send them back to the original network traffic, as shown in figure~\ref{fig:1}. The PACS server receives the malicious DICOM and sends these files to the radiologist's workstation, eventually infecting the system. 

\subsection{Infiltrating Malicious Payload in DICOM}
\label{infiltrating dicom}

As the attacker now has access to the DICOM files, it can easily get manipulated by the malicious payload. The following two sections describe the vulnerability in the DICOM files and the proposed malware-injecting method.

\subsubsection{Vulnerability in DICOM}
\label{vulnerability in dicom}

\begin{figure}[ht]
\centering
\includegraphics[width=0.4\textwidth]{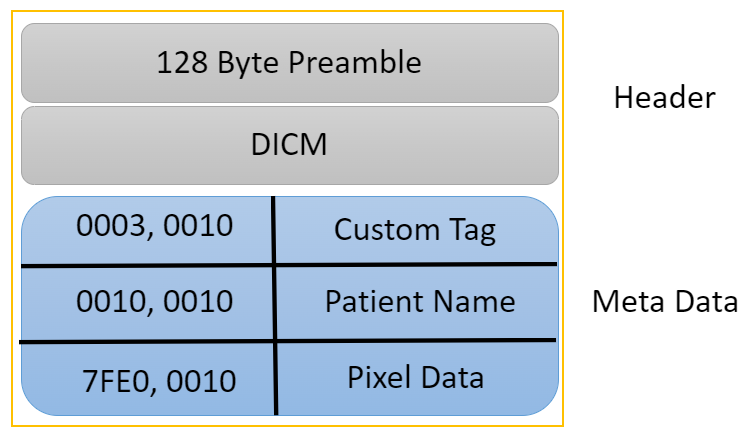}
\caption{DICOM File Structure}
\label{fig:2}
\end{figure}

A DICOM file contains three essential sections: the header, public and private tags, as depicted in Figure \ref{fig:2}. The header section of the DICOM has a 128-Byte preamble. The next four bytes with the prefix "DICM" is the magic number with 4 Bytes. The public tags are the even-numbered tags that have a specific meaning associated with them. For example, the tag 0010 is always associated with the patient's name. The private tags are the odd-numbered tags that have the freedom to add their custom tag.

All hospitals support different formats of images like TIFF, JPEG, JPG, and NIFTI. To make every DICOM support all file formats, the headers of these formats can easily be inserted in the empty preamble section and then create a pixel pointer to the pixel data tag. This way, the .dcm format can smoothly transition from other file formats. However, this empty preamble section exposes vulnerabilities in the DICOM files and makes an excellent target for attackers by inserting malicious code.

\begin{figure*}[ht]
\centering
\includegraphics[scale=0.5]{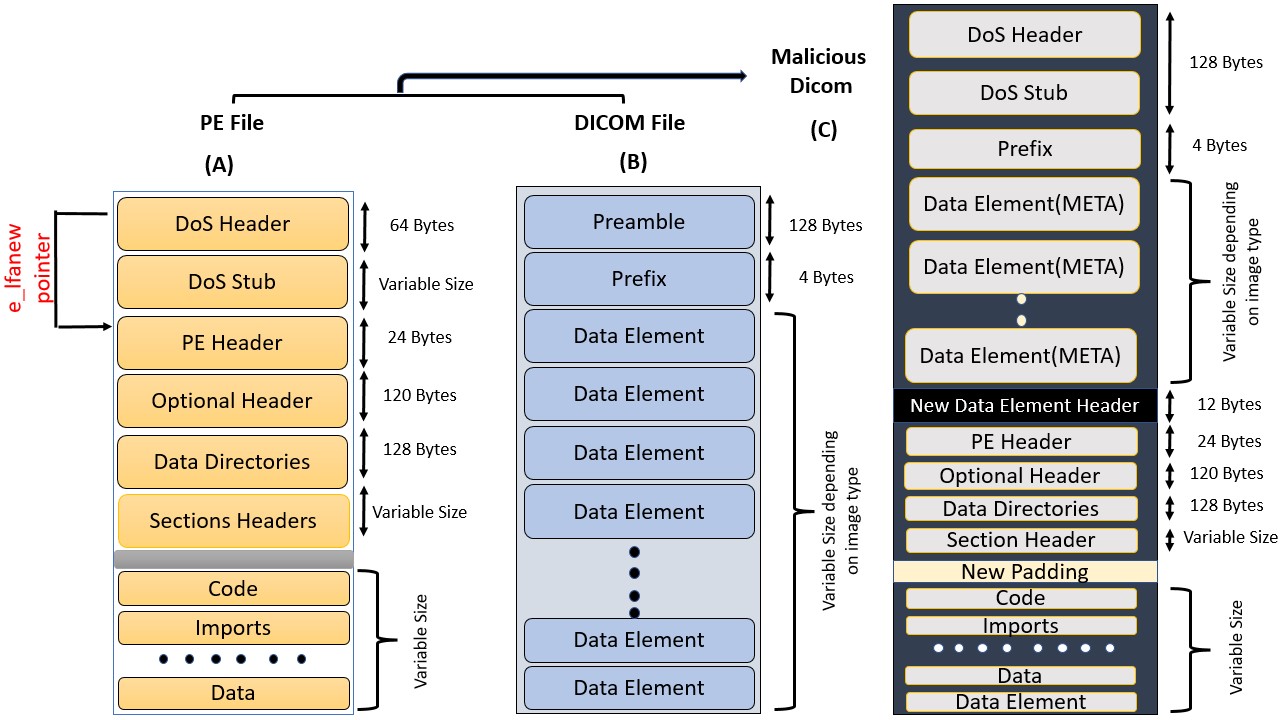}
\caption{Infiltrating Malware Executable into DICOM Files. Part (A) is the PE file of a Windows malware, where the DoS header of 64 Bytes points to the PE header using an e\_lfanew pointer. Part (B) is the structure of the DICOM file, where the preamble section of 128 Bytes is the vulnerability where the DoS Header and DoS stub from the PE file are injected. Part (C) is the malicious DICOM file, with the DoS header and Stub in the preamble and the remaining PE in the private tag of DICOM. The e\_lfanew pointer is updated to point to the PE header with variable size DoS Stub.}
\label{fig:3}
\end{figure*}

An adversary can inject malware in the empty preamble header section of the DICOM file. There have been many techniques to insert malware inside images, like steganography\cite{stegano}\cite{dicom_risks}, vector embedding\cite{vector}, and statistical techniques\cite{statistical}. However, these techniques are built to hide the malware and not affect the application running for viewing the DICOM. Our proposed attack MalDicom aims to turn a good process into a bad one. The targeted process is the DICOM viewer\cite{dicom_viewer} in the radiologist's workstation. In this paper, we perform code injection by infiltrating byte-by-byte malware into the preamble section of DICOM. Injecting a malicious payload often evades detection. The injection process will make the sandbox unaware of the malware. An antivirus program installed in the system will also bypass the malware. The CT or MRI scan is still in the correct DICOM format but contains malware. It is capable of destroying the victim's machine with a single click. The malware code injection into a DICOM file is demonstrated in the following subsection.  

\subsubsection{Process of Code Injection into DICOM Files }
\label{code injection}

\begin{figure*}[ht]
\centering
\includegraphics[scale=0.5]{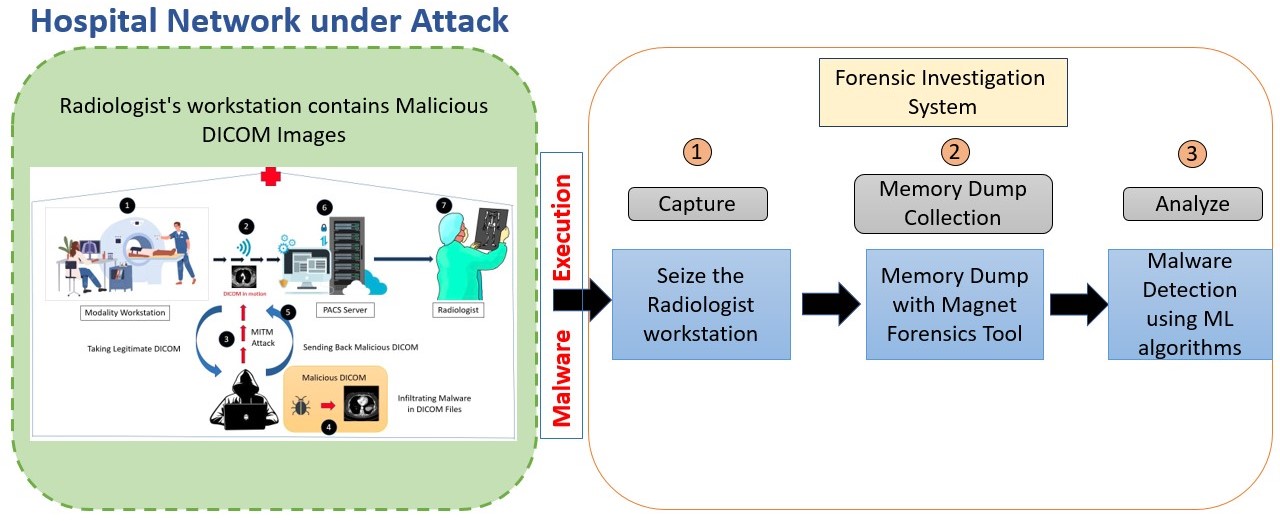}
\caption{A Memory Forensics Investigation Framework: MalDicom }
\label{fig:5}
\end{figure*}

\begin{algorithm}
\caption{Code Injection of Malicious payload into DICOM file}\label{alg:cap}
\begin{algorithmic}[1]
\Function{$CreatePEDicom$}{pePath, dicomPath, malDicomPath}: 
\State $SnipDoSFromPE(pePath, sizeofDicomMetaData)$
\State Call the $e\_lfanew$ pointer
\State$DoS\_Stub\_Size=DoS\_Header\_Size-PE\_Header$
\State$variable\_length=DOS\_Header['e\_lfanew']['Value']$ 
\State Fill the space between DoS Header and the PE header by $sizeofDicomMetaData$
\State $ e\_lfanew=updated $
\State $ variable\_DoS=DoS\_Header\_Size+DoS\_Stub\_Size+DICOM\_Magic\_Number+sizeofDicomMetaData +PE\_Header $
\State $ DOS\_Header.e\_lfanew=variable\_length+4+sizeofDicomMetaData+24 $
\State $DICOM.preamble=variable\_DoS$
\State Generate an Intermediate DICOM file
\State $ remaining\_PE\_file=SnipRemainingFromPE $  
\State $ padding=bytes(paddingNeeded)$
\State $ pe\_header+=padding $ 
\State $ pe\_header+=remaining\_pe $
\State Add the remaining PE to the DICOM file
\EndFunction
\end{algorithmic}
\end{algorithm}

For injecting the malware into the DICOM, we consider Windows malware executable. Figure \ref{fig:3} shows the infiltration of malware exe file into DICOM. Part (A) in figure~\ref{fig:3} is the structure of the PE executable file. The exe file contains the DoS Header with 64 Bytes and the PE header with 24 Bytes. An $e\_lfanew$ pointer points from the DoS header to the PE header. The difference between the two gives the size of the DoS Stub. (B) in figure~\ref{fig:3} shows the DICOM file's structure, which we discussed in the previous section [Figure~\ref{fig:2}]. Part (C) in figure~\ref{fig:3} is the malicious DICOM file with the exe file infiltrated into the DICOM. Algorithm~\ref{alg:cap} shows steps to inject a malicious payload into DICOM. It takes three arguments: the malware EXE file, the DICOM file in which we want to insert the malware, and the resultant DICOM. A $CreatePEDicom$ function was created and was called with these three arguments. The $SnipDoSFromPE$ is called to cut the DoS Header and Stub from PE. It takes the PE path and metadata of the DICOM as arguments. Now, the DoS header contains the $e\_lfanew$ pointer, which points to the PE header. Therefore, the size of the DoS stub will be the difference between both. During the insertion of the DoS header and the DoS stub in the preamble section, the DICOM metadata will also be available between them as shown in figure ~\ref{fig:3} part C. After that, the $e\_lfanew$ pointer will be updated. We call this file Intermediate DICOM. The remaining section of the PE is then inserted in the private tag of the DICOM. During the insertion of the remaining PE file, a few extra bytes are created that can be compensated by adding padding after the header section. Finally, the remaining PE is added to the DICOM file. This way, the malicious payload can be inserted into the DICOM file without modifying its content.

The infected DICOM's evaluation and memory forensic investigation has been performed using an open-source memory dump dataset, CIC-MalMem\cite{cic}, as explained in section~\ref{investigation}. We took the DICOM samples of Lung CT scan from Cancer Imaging Archive website\cite{image}. The .exe file of the Trojan virus performs code injection attacks on the Lung CT scan DICOM files.



\section{Investigation Process}
\label{investigation}
Figure~\ref{fig:5} shows the proposed framework, MalDicom, for the memory forensic investigation. The forensic investigation system contains three crucial steps:

\subsubsection{Seize the Radiologist's Workstation}: The first step is to seize the system for forensic investigation to avoid evidence tampering. 
\subsubsection{Memory Dump Collection}: We took the memory dump of the machine using Magnet forensics\cite{magnet}, a Windows-based forensics tool. 
\subsubsection{Analysis using ML Algorithms}: Once we have the memory dump, different ML algorithms can be used to investigate the malware as described in the next subsection~\ref{detection} 

\subsection{Dataset}
\label{dataset}

For forensic investigation, we consider the CIC-MalMem\cite{cic} dataset 2022, which has a collection of different categories of the malware memory dump. It contains benign and malignant malware samples to investigate the infected system. The categories of malware memory dumps include Spyware, Trojan Horse, and Ransomware. The categories contains fifteen families of malware, five from each type of malware. The dataset contains 58,596 records, with 29,298 benign and 29,298 malicious memory dumps.

\subsection{Feature Importance}
\label{feature}

Feature importance assists in estimating the contribution of each feature in the dataset to the model's prediction. Determining which features impact the model's prediction of essential decision-making is crucial. In memory forensics, this can help determine which features a forensic investigator should focus on more to detect adversaries faster. T
To analytically evaluate the contribution of each feature to the ML model, we have used Shapley values\cite{shapley}. It is a method from coalition game theory that describes allocating the outcome among features equitably. It can be formulated in the following way:\\
There are $N$ malware samples with malware as 1,2,3..., $N$, and a function $M$ that takes the subset of malware samples and returns the real-valued outcome in the malware detection. Formally, the contribution of malware sample $i$ is defined as:\\
\begin{equation}
    \phi (M)= \sum_{C \subset N   {i} } \frac {|C|! (N-|C| -1)!}{N!} (v(C \cup {i}) - M(C))
\end{equation}

\begin{figure}[ht]
\centering
\includegraphics[scale=0.5]{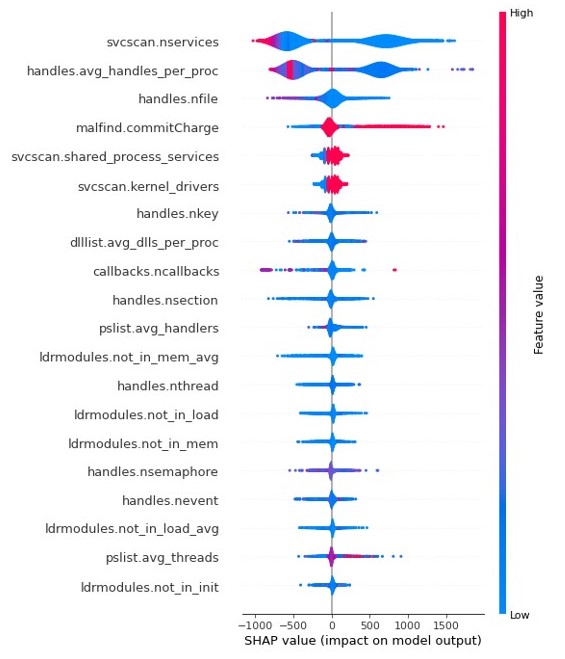}
\caption{Shapley Values}
\label{fig:6}
\end{figure}

Where C is a coalition, or subset, of malware. $M(C)$ is the prediction for feature values in set C that are marginalized over features that are not included in set S:
\begin{equation}
    M(C)= \int\limits \widehat{f} (m1, m2, ...., m_n)d \mathbb{P}_{m\epsilon C} - E_m(\widehat{f}(M))
\end{equation}
Here, $f$ is the feature value for malware subset $C$, and $E_m$ is the effective estimate for feature values. The Shapley value permits contrastive explanations based on valid theory and fairly distributes effects. Figure~\ref{fig:6} shows the Shapley values of the various features for malware memory dump present in the dataset, CIC-MalMem. It shows that feature $malfindCommitCharge$ has a high shap value. This feature is essential to detect malware using the ML model. The $malfind$\cite{malfind} command is used in memory forensic investigation to help find the injected code or dynamic linked list (DLLs) in memory. Therefore, shap values explain the $malfind$ feature importance in ML model output prediction. Next, we can detect the malware families using ML algorithms described in the following section.

\subsection{Detection using Machine Learning Algorithms}
\label{detection}
A total of fifty-seven features, including $pslist$, $dlllidt$, $handles$, $ldrmodules$, $malfind$, $psxview$, $svcscan$, and $callbacks$ collected by the Volatility tool, are used for memory forensic investigation. The malware families and benign families contain unbalanced samples. Thus, we use the SMOTE\cite{smote} algorithm to over-sample the minority samples. e perform multi-class classification for malware detection. The detection was evaluated on six classifiers: Decision Trees (DT), Support Vector Machine (SVM), Gaussian Naive Bayes (NB), K-nearest Neighbors (KNN), Random Forest (RF), and Multi-Layer Perceptron (MLP). The dataset was divided into an 80:20 ratio for training and testing, respectively.

\section{Results}
\label{results1}
We evaluated multiple ML models for multi-class classification of malware detection. Table~\ref{table4} shows the results of the ML models trained on memory dumps from the CIC-MalMem dataset to predict which category each malware family belongs to. The RF model outperforms all the other models with an accuracy of 75\%. Next, we have the DT model with an accuracy of 72\%. The other three models performed in an average manner with 50\% accuracy. The F1 score for the top 3 classifiers were 0.53, 0.48, and 0.36. We also evaluated the results on each malware family for the top three ML classifiers. The malware memory dump samples that were easily tracked were Trojan Refroso and Spyware-TIBS. The RF model for Trojan Refroso gives an F1-score of 0.75, and for Spyware-TIBS is 0.62. The DT model for both dumps had a score of 0.68 and 0.65. The KNN classifier scored 0.53 for Trojan Refroso and 0.42 for Spyware-TIBS. The Ransomware families gave less score as compared to other malware families.

\begin{figure}[ht]
\centering
\includegraphics[scale=0.5]{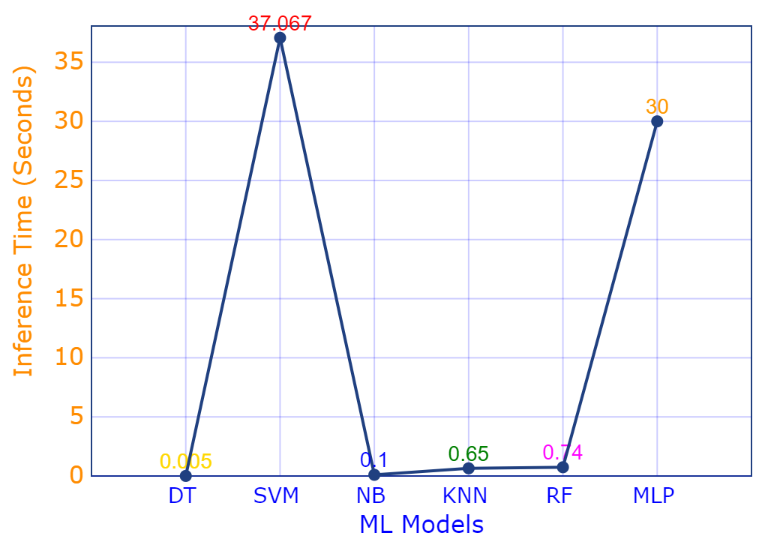}
\caption{Inference Time of ML Models}
\label{fig:7}
\end{figure}

All the classifiers were finally evaluated on unseen 20\% data from the dataset. We calculated inference times to determine the ML model with the lowest detection latency. Figure~\ref{fig:7} shows the inference time of all the models. The observation shows that the DT was the fastest of all. After that, Naive Bayes and the KNN are the second-best models. The SVM classifier takes the highest time for real-time malware detection. The MLP model took 30 seconds, comparatively more than other ML classifiers.



\begin{table}
\centering
\caption{Classification Result}
\label{table4}
\begin{tabular}{|p{1.75cm}|p{1.25cm}|p{1.2cm}|p{1cm}|p{1.3cm}|}
\hline
\textbf{Model}&\textbf{Accuracy} &\textbf{Precision} &\textbf{Recall}&\textbf{F1-Score}\\
\hline
DT & \textbf{0.72}& 0.48& 0.48& \textbf{0.48}\\
\hline
SVM &0.56& 0.40& 0.19& 0.17\\
\hline
Gaussian NB &0.56&0.25&0.20&0.15\\
\hline
KNN Classifier &\textbf{0.66}&0.37& 0.37& \textbf{0.36}\\
\hline
RF &\textbf{0.75}& 0.53& 0.53& \textbf{0.53}\\
\hline
MLP & 0.50& 0.20& 0.15& 0.10\\
\hline
\end{tabular}
\end{table}



\section{Discussion and Future Work}
\label{future work}
The paper presents a novel attack methodology for compromising the hospital network through malicious DICOM files. After inserting the malware in the DICOM file, the memory dump of the affected system was later analyzed using ML algorithms. However, detecting the malicious memory of DICOM using ML algorithms is a time-consuming and computationally expensive task. Therefore, to address this challenge in the future, malware detection can be done using entropy. Entropy is a measure of randomness. Shanon's entropy\cite{shanon} can be used to find noisy data in the preamble section of the DICOM file. The entropy method is encouraged because scanning the DICOM file takes constant time, O(1). The overhead will be negligible using entropy, and we can easily detect the malicious DICOM file.  



\section{Conclusion}
\label{conclusion}
The ever-growing network of connected medical devices has created a massive internal blind spot where cyber-attacks are prominent. This paper introduces a novel attack technique on DICOM files using code injection. We demonstrate how skillfully an adversary can take the DICOM data, insert malware and send back the malicious DICOM pretending to be legitimate DICOM. On viewing the compromised DICOM file, the hidden malware gets executed on the system and infects its memory. We proposed a memory forensics framework, MalDicom, to accelerate the investigation process and analyze the infected memory to classify the underline malware. We used the CIC-MalMem dataset to retrieve various malware families' memories. The memory dump was then analyzed using six different ML algorithms and got an accuracy of 75\% with the Random Forest algorithm. For the ML model explainability, we used the Shapley values to identify the impact of ML input features on output predictions. The feature $malfindCommitCharge$ has received the highest Shapley values for the CIC-MalMem dataset. It infers the presence of malware in the ML input data.  



\section*{Acknowledgment}
We acknowledge Mr. Shyam Sundar Ramaswami from GE Healthcare, India. We are grateful for his insights into DICOM security issues.

\bibliographystyle{IEEEtran}
\bibliography{conference_101719}

\end{document}